# Modern analog computing for solving differential and matrix equations

Zhong Sun[1*], Piergiulio Mannocci[2], Manuel Le Gallo[3] & Abu Sebastian[3]

[1]Institute for Artificial Intelligence, School of Integrated Circuits, Peking University, Beijing Advanced Innovation Center for Integrated Circuits, Beijing 100871, China.

[2]Dipartimento di Elettronica, Informazione e Bioingegneria, Politecnico di Milano, Piazza Leonardo da Vinci 32, Milano, 20133, Italy.

[3]IBM Research Europe, Rüschlikon, Switzerland.

*e-mail: zhong.sun@pku.edu.cn

In recent years, driven by the computational demands of data-intensive applications such as artificial intelligence and scientific computing, analog computing has gained renewed interest. Given the diversity of computational tasks and recent advancements in analog CMOS circuits and resistive memory technologies, we refer to the evolving landscape as modern analog computing. In this context, we identify three core computational primitives: solving differential equations, solving matrix equations, and performing matrix-vector multiplications, and we explore the connections among them. We also examine various hardware implementations of these analog computing operators, including those built with discrete components, integrated circuits, and resistive memory devices. Among these, resistive memory arrays emerge as particularly promising due to their implementation efficiency. The paper then surveys recent progress in leveraging modern analog computing to solve differential and matrix equations using both advanced analog CMOS circuits and resistive memory arrays. Finally, we discuss the applications of these circuits, the precision and scalability issues and their potential solutions, the relationship with in-memory computing, and the unique computational complexity of analog computing. This paper provides a unified perspective on analog computing, highlighting its strengths, current developments, and challenges, and positioning it as a pivotal enabler of next-generation computational frontiers.

## Introduction

One who has already become familiar with experience in the physical world may find that the digital computer we are using is sort of counter-intuitive. It is a highly abstract symbolic system that performs general-purpose computing based on the Turing machine model and Boolean algebra. While acknowledging its overwhelming dominance of use in modern society, it is reasonable to complain about the resulting high complexity of transistor connections in the primitive circuits, which hinder a clear understanding of the computing process. For instance, adding 3 single bits (which is a 1-bit full adder) requires a circuit composed of 28 transistors, but the physical laws of nature are continuously performing addition operations with no design, in the form of current, charge, or whatever can be accumulated[1]. Such a contrast will be accentuated if one considers the addition or multiplication of two real numbers that are represented as many bits. Additionally, for solving any problem, elaborated algorithms should be designed by performing a sequence of atomic operations such as multiply-accumulate (MAC), where the hardware cost and time consumption are described by the concept of computational complexity. Today, due to the rapid development of artificial intelligence, scientific computing and other many data-intensive applications, such a complexity creates intractable workloads for computers[2].

By contrast, the principle of an analog computer is very closely in line with human intuition for solving problems. That is, analogizing the mathematical equation to a reconfigurable physical (mechanical, electrical, electronic or biological) system, the operation of which outputs results analogy to the solution[3,4]. It is for this reason that it became the early form of the computer, whether it was the Antikythera machine in Ancient Greek BC or the mechanical and electrical analog computer at the beginning of modern times[5-8]. In an analog computer, achieving massive computational parallelism is straightforward due to its hardware architecture. Alternatively, it can be viewed as a parallel computing algorithm (or sub-algorithm) embedded within the hardware system[9]. Since the 1960s, driven by the powerful Moore’s Law, digital computers have continued to develop rapidly for decades, while the development of analog computing has lagged behind. Only recently, because of the decline of Moore’s law and the existing von Neumann bottleneck in digital computers, analog computing has attracted renewed attention[10].

It is equipped with new concepts of both hardware device and computing methodology, in pursuit of continuous improvement and breakthrough of computer performance. By directly using the inherent physical laws of the devices and circuits in the chip to perform computations, the high spatial parallelism in the hardware architecture results in fast speed and low complexity that is very beneficial for big data processing[11].

There are currently two interpretations of analog computing. One views analog and digital computing as opposites, where analog involves continuous signals, and digital operates with discrete values. The other emphasizes the analogy between mathematical problems and physical systems without restricting model variables and physical quantities to be continuous or discrete[12]. We prefer the second definition, which better captures the essence of analog computing. In addition, the view that analog computing is continuous is just a by-product of the second view, because analog computing focus more on solving continuous math problems, which are represented by variables such as voltage and current. However, it may also be used to implement discontinuous functions, such as the Heaviside step function in a perceptron network, again by using analogous physical phenomena. A perceptron network can be used as Boolean logic gates by the concept of threshold logic, and eventually for general-purpose digital computing[13]. Nevertheless, the principle is substantially different from the existing approach based on CMOS logic gates.

In this work, we present a comprehensive review of modern analog computing for solving differential equations and matrix equations. We begin by introducing the core analog computing primitives, including solving differential equation, solving matrix equation, and performing matrix multiplication, all of which are composed of the same set of basic operators. We also highlight their inherent interconnections. We then explore various hardware implementations of these operators, including CMOS analog circuits and resistive memory arrays, as well as circuit designs for solving differential and matrix equations. Special attention is given to resistive memory arrays, which represent a highly compact integration of analog operators and enable efficient in-memory analog matrix multiplication and matrix equation solving. Furthermore, we discuss the challenges of precision and scalability in analog computing, highlighting hybrid analog-digital approaches that enhance accuracy and extend the computational capabilities of analog circuits.

## Analog computing primitives and operators

Analog computers are generally used to solve specific problems which are extensively involved in important applications. Traditional analog computing focuses primarily on solving differential equations, including ordinary differential equations (ODEs) and partial differential equations (PDEs)[14-16]. In modern times, the critical role of matrix computation across diverse applications, coupled with advancements in emerging device technologies such as resistive switching memories, has reignited interest in leveraging analog computing principles for efficient matrix multiplication[17-19] and matrix equation solving[20-22]. The three types of computing primitives are summarized in **Fig. 1**. They are independent of each other and can be used to accelerate computations in specific applications. At the same time, there is a close relationship between the three primitives.

Specifically, when solving PDEs, space discretization is required to approximate partial derivatives as weighted combinations of variable values at neighboring grid points. If the equation is time-independent, it is transformed into a system of linear algebraic equations, forming a matrix equation, where numerical coefficients represent linearized operators acting on neighboring grid points. If the equation is time-dependent, the PDE is converted into a system of ODEs[23]. ODEs can be solved directly by designing analog circuits, but can also be discretized into matrix equations over time, which are in turn solved by analog circuits, resulting in solutions at discrete spatial and temporal grid points. Matrix equations can also be solved by iterative algorithms[24], which typically involve matrix multiplications combined with vector or scalar operations, and then analog circuits are used to accelerate the core matrix multiplication operations. Therefore, there are multiple paths for analog implementation of these three types of computing tasks, allowing for diverse analog computing designs and applications. In contrast, digital computers can only follow a single path through matrix multiplication, matrix equation solving, and differential equation solving, therefore lacking the flexibility of implementation strategies.

Although computing tasks fall into three distinct categories, their designs rely on the same basic operators. In the 1930s, Shannon demonstrated that a set of basic operators—adder, multiplier, integrator, and constant input (**Fig. 2a**)—constitutes a functionally complete system for general-purpose analog computing, particularly for solving differential equations[25,26]. These operators are also essential for matrix multiplication and matrix equation solving. Note that, the constant input term is often ignored in modern designs and can be conveniently provided by a digital-to-analog converter (DAC)[27,28]. The circuit topologies for these three computing primitives are illustrated in **Fig. 2b-2d**, respectively. For broader applications, such as solving nonlinear matrix equations, additional nonlinear function operators are required[23]. Most analog computing approaches operate in the voltage domain, where a voltage signal can naturally fan out to multiple branches. Consequently, fan-out operation is typically disregarded in voltage-based implementations. However, this is not the case for current-based analog computing[29], as it will be discussed later.

The implementation of matrix multiplication requires only multipliers and adders and is realized in an open-loop configuration without integrators or feedback (**Fig. 2b**). In contrast, differential equation solving and matrix equation solving involve adders, multipliers, and integrators (**Fig. 2c** and **2d**). The integrator accounts for the differential operation in the equation, and its output will be fed back to the input of the circuit to form a closed loop, thus solving the equation in continuous time. While both rely on similar components, their output results are different. In differential equation solving, the evolution of the circuit's output over time is concerned, and the result consists of the output at each time point (determined by the sampling frequency). Matrix equation solving in analog computing is essentially a process of solving a differential matrix equation (or a set of ODEs), but only the circuit's stable output is concerned, instead of its dynamic evolution. **Fig. 2d** illustrates the analog computing circuit for solving a system of linear equations (a matrix inversion problem) $\boldsymbol{Ax} = \boldsymbol{b}$, whose dynamic process is described by a differential equation, that is, $d\boldsymbol{x}/dt = \boldsymbol{Ax} - \boldsymbol{b}$. This process can be interpreted as continuous-time gradient descent, where the solution vector $\boldsymbol{x}(t)$ converges as $d\boldsymbol{x}/dt$ approaches zero, yielding the matrix equation solution at steady state[30]. More generally, solving various matrix equations in analog computing follows the same principle:

implementing a system of differential equations and measuring the circuit's steady-state output. This highlights the deep connection between differential equation solving and matrix equation solving, as differential equations can be solved by discretizing variables into a matrix equation, while matrix equation solving in analog computing fundamentally reduces to solving a system of ODEs with convergent steady-state outputs.

Analog computing operates through fixed topological connections between fundamental operators, with its implementation dependent on both the design of hardware operators and the choice of physical variables. In addition to adder, multiplier, and integrator, fan-out operation also requires a dedicated hardware module where current is used as the physical variable. As device and circuit technologies have advanced, the hardware implementation of these operators has evolved, as summarized in **Fig. 3**. Early analog computers implemented multipliers using discrete potentiometers, requiring hundreds or thousands of such components to provide programmable parameters[31,32]. Their operation is based on Ohm's law, where one operand is fixed after programming while the other is an external voltage. With the advent of integrated circuits (ICs), CMOS amplifier-based multipliers[1,33-35], such as the Gilbert cell, became the primary approach. These circuits allowed adjustable parameters for multiplying two input signals (voltage, current, or both) or performing constant-variable and variable-variable multiplications. However, both methods had their own drawbacks: potentiometers were not integrable components, and CMOS amplifier circuits had complex structures with high resource consumption, limiting the scalability of analog computing.

The emergence of resistive switching memory offers a promising alternative. This non-volatile, adjustable resistive device shares the same fundamental computing principle as potentiometers but features a simpler structure and better integrability, enabling large-scale array implementations[36-40]. Compared to CMOS circuits, resistive switching memory significantly reduces area and power consumption. For example, at the 65 nm technology node, a typical CMOS multiplier consumes tens of microwatts (μW) and occupies hundreds of square microns ($\mu m^2$)[30,41], whereas resistive memory devices achieve microwatt-level power consumption with an area of just hundreds of square nanometers ($nm^2$)[42,43]. This translates to orders-of-magnitude improvements in integration density and energy efficiency, greatly enhancing the scalability of analog computing circuits.

In traditional analog computers, adders are primarily implemented using operational amplifiers (OPAs), operating in the voltage domain, where both inputs and outputs are voltages. However, if current is used as the physical variable, addition is inherently free, as it can be performed directly via Kirchhoff's current law (KCL). The situation is reversed for fan-out: in voltage-domain computing, fan-out occurs naturally, whereas in current-domain computing, the design of fan-out is mandatory, to make copies of the current signal[41,44]. In IC designs, current mirrors are commonly employed for current fan-out within a chip, distributing a single reference current to multiple output branches serving different functional blocks. To achieve this, a conventional single-output current mirror can be extended to multiple outputs, provided that the power supply and reference current source can sustain the increased total current without inducing degradation or instability. Additionally, sufficient voltage headroom must be maintained across all branches to prevent transistors from entering the linear (triode) region, which would degrade current mirroring accuracy.

Given that both addition and fan-out can be implemented without extra hardware in different domains, an optimal design combines the advantages of both—using voltage inputs for free fan-out and current outputs for free addition. When paired with resistive memory device as a multiplier, this approach naturally leads to the resistive memory crossbar array, which offers the most compact integration of all three basic operations. The crossbar array is an ideal structure for maximizing computational density, leveraging the high regularity of matrix operations for efficient parallel computing[45,46]. In the array, applying voltages to the columns while grounding the rows enables instantaneous, highly parallel matrix-vector multiplication (MVM) through direct current accumulation. Therefore, it is highly promising for large-scale analog computing applications. Additionally, performing computations *in situ* within resistive memory helps overcome the von Neumann bottleneck, further enhancing computational performance and energy efficiency[13,47].

The core component of an integrator is always a capacitor, which accumulates charge over time,

producing a voltage as a result[44,48,49]. As variables could be positive or negative, the differential design can be used, where the positive and negative input currents are integrated in the two plates of the capacitor, respectively. Integrators are usually implemented with active circuits, most commonly amplifier-based designs. An OPA-based integrator functions similarly to an adder but replaces the feedback resistor with a capacitor. In ICs, integrators are often designed using operational transconductance amplifiers (OTAs), which offer a more compact architecture and can be seamlessly integrated with other computational modules on the same chip. For the conversion between current and voltage, transconductance amplifiers and transimpedance amplifiers (TIAs) should be used. In addition, an initial condition setting block imposes predefined initial voltages on the integration capacitor, establishing specific starting points for the integration process[41]. On the other hand, recent research has shown that a single-pole OPA can function as an integrator in analog computing with resistive memory arrays[50-52]. As a high-gain voltage-controlled voltage source, the OPA's frequency response is governed by its dominant pole, set by an equivalent RC network that defines the associated time constant. This approach can be efficiently implemented using ICs, ensuring seamless compatibility with large-scale analog computing systems.

In principle, various hardware operators can be combined to construct an analog computing circuit, in either voltage or current domain, or the hybrid domain in resistive arrays, for performing the three types of computing tasks. As such, there should be many approaches towards the implementation of an analog computer. Based on the observation of modern IC developments and emerging resistive memory technology, three mainstream approaches are described in the following sections.

**Solving equations with advanced CMOS analog circuits**

Over the past few decades, while digital circuit performance has improved exponentially following Moore's Law, analog circuit technology has also made significant strides, albeit at a slower pace[9]. Consequently, there have been ongoing efforts to leverage advances in analog IC technology to develop high-performance analog computing systems[9]. A prototype analog CMOS accelerator chip was fabricated at the 65 nm technology node[29,41] (**Fig. 4a**), which was a successor to an earlier design fabricated in a 0.25 μm CMOS process, with upgraded functions[28,44]. The chip uses current to encode data, so the adder can be achieved without a design, but the implementation of fan-out becomes necessary and is achieved by current mirror that fans out one input into three outputs. The multiplier design utilizes a network of MOSFETs operating in the weak inversion region, leveraging their exponential I-V characteristics for precise analog multiplication. A signal stored in the circuit's memory controls the multiplier, allowing it to function either as a 2-input, 4-quadrant multiplier or as a constant multiplier in a variable gain amplifier (VGA) configuration. To ensure accuracy and stability, the integrator design is relatively complex. It consists of five main components: a capacitor, two input current mirror circuits, a common-mode feedback block, two output transconductors, and an initial condition setting block.

For functional testing, the chip includes a limited set of operator blocks, comprising four integrators, eight multipliers, and eight fan-out operators. These blocks are organized in an array and interconnected via a reconfigurable crossbar switch network. A serial peripheral interface (SPI) is used to program the chip, enabling mode selection for various operations, including configuring signal paths, setting block parameters, executing computations with the configured circuit, and calibrating analog imperfections. Additionally, the SPI facilitates interfacing between analog and digital signals. To address nonlinear differential equations, a SRAM module is designed with a continuous-time digital data path. It allows for the implementation of arbitrary nonlinear functions as a linear lookup table (with a precision of 8-bit), which is much more advantageous than the hardwired few specific types of nonlinear functions[44]. In addition, the chip contains a set of on-chip DACs and analog-to-digital converters (ADCs), which are also designed as a continuous-time current mode, thus allowing the interaction with continuous-time SRAM.

This analog chip has been used to solve several canonical differential equations. One example is the van der Pol equation, which is a second-order nonlinear differential equation that describes a non-conservative, oscillating system with non-linear damping[41]. **Fig. 4b** presents the equation form and the corresponding circuit block diagram. A chain of two integrators is constructed to convey the integration from the second-order differentiation of the variable $x$. Since the equation includes a quadratic term, the multiplier operates as a 2-input multiplier to generate the $x^2$ term. Additionally, it functions as a VGA to perform constant scaling. Given that multiple signals are reused throughout the circuit, several fan-out modules are incorporated. Notably, a single current mirror provides only three outputs, which may be insufficient. This limitation is addressed by cascading two current mirrors to extend the number of available outputs. The real-time solution of the equation, shown in **Fig. 4c**, exhibits the expected oscillatory behavior. The output signals must be continuously sampled to monitor the oscillation dynamics. The circuit results are compared with the ideal numerical solution, yielding a relative mean square (RMS) error of 4.6% over a 60-second period. This low error is achieved through digitally-assisted calibration, which compensates for non-idealities in all analog blocks, including DC offset biases at the integrator inputs and outputs, multiplier gain errors, and nonlinearities in DC transfer characteristics. The van der Pol oscillator simulation is completed in only 480 μs with an energy consumption of 0.14 μJ, providing approximately two orders of magnitude improvement in both solution time and energy efficiency compared with a digital processor[41].

This chip has been successfully used to solve arbitrary nonlinear differential equations by utilizing an SRAM-based nonlinear function generator operating in continuous time, together with continuous-time ADCs and DACs. For example, it has been applied to emulate a coupled mass-spring system with nonlinear springs, involving operations such as absolute value, sign, and square root[41]. The chip solves the nonlinear differential equation in 320 μs, enabling the observation of the mass-spring system dynamics over a 40-second timescale, with an RMS error of 4.7% and an energy consumption of 0.25 μJ. These results highlight the efficiency of analog computing for low-power scientific computing applications. Various linear and nonlinear and differential equations have been solved using an analog compilation toolchain specifically developed for this chip[27,53]. For solving PDEs, such as Maxwell's Equations, the spatial domain is first discretized using finite-difference methods, after which the resulting continuous-time ODEs are solved using analog circuits to obtain the solution at each spatial point. Although this approach has been extensively explored in OPA-based analog computing[54-

[56], it has not yet been implemented in integrated analog chips until recently.

With the same chip, the matrix equation ***Ax*** = ***b*** can also be solved, although it was not designed primarily as a matrix equation solver. As the solution to this matrix equation is the equilibrium output of a system of first-order differential equation, the circuit can be configured similarly as the differential equation solving case[30,57]. In this system, there are multiple individual differential equations that interact with each other, which are integrated in parallel, and the output is fed back to all branches. As a result, the connection becomes much more complicated than the single high-order differential equation solving, as shown in **Fig. 4d**. That said, the system could be much clearer when presented in the matrix form in **Fig. 2d**. In this system, all multipliers are used as VGAs that encode all entries of matrix -***A***. The known vector ***b*** is provided by a set of DACs. Limited by the number of output terminals of the fan-out module, only a 2×2 matrix can be processed, and the remaining output of each fan-out is used for readout by ADC. **Fig. 4e** shows an example of the transient output of this configuration circuit, which reaches an equilibrium state within a certain time. In this case, only the stable output should be sampled, which releases the requirement on the sampling rate and the accompanying cost. Due to the stability issue caused by the complicated feedback connections, the concerned matrix ***A*** is constrained to be positive definitive. If this condition is not satisfied, the output will diverge and eventually be limited by the maximal available signal range. A similar design has been proposed, by using a network of transconductance amplifiers and a set of capacitors based on the FPAA platform[58], where the condition on positive definiteness also applies. To overcome this limitation, two sets of matrix coefficient blocks (together with one additional set of amplifiers) can be used to construct a global feedback circuit. However, this is only applicable in voltage-mode analog circuits and has been implemented with OPAs[59]. Based on this principle, the chip can be used to solve more problems, including nonlinear PDEs, such as the viscous Burgers' equation[23] and the 2D Poisson elliptic PDE[30]. By discretizing the PDEs in both space and time into a nonlinear system of equations, the matrix equation solving circuit can be used to solve the resulting Jacobian matrix inverse problems according to Newton's method.

**Solving matrix equations with resistive memory circuits**

Over the past two decades, the physical property of resistance (or conductance) has been explored for information storage and processing[36]. This has led to the development of various resistive memory device concepts[47,60,61], including resistive random-access memory (RRAM), phase change memory (PCM), magneto-resistive memory (MRAM), ferroelectric tunnel junctions (FTJ), and ionic (synaptic) transistors, among others. While these devices differ in their underlying physical mechanisms and structures, they share a common feature: the conductance state of the device can be controlled by external voltage or current stimuli and stored in a nonvolatile manner[13]. These devices are sometimes discussed within the unified framework of memristors[62] and can typically be tuned to multiple levels (or even quasi-continuous states), as illustrated in **Fig. 5a**. Although passive 1R arrays can be used for memory and computing applications, the more mature 1T1R structure is preferred for reliable operation. A crosspoint array can be used to represent a positive matrix for analog computing. Prior to computation, each resistive memory device must be pre-set, typically through regular write operations. Analog conductance tuning is usually achieved using write-verify schemes, generally requiring several tens of write cycles to reach a predefined conductance state with an error of only a few percent[63-65]. In this context, the resistive memory device functions as a constant multiplier, unlike CMOS-based devices, and therefore cannot be used as a 2-input multiplier.

To construct analog circuits for solving matrix equations, crosspoint resistive memory arrays are combined with OPAs that are used as integrators. Unlike traditional analog computing, which relies on separate adders, multipliers, integrators, and fan-out circuits as individual building blocks for solving differential equations, this approach utilizes the crosspoint arrays and OPAs as the core building blocks. Building on this straightforward principle, a series of analog matrix computing (AMC) circuits have been designed to efficiently solve various matrix equations[11]. Although this approach can be applied to differential equations after their discretization into matrix equations, the AMC circuits themselves are broadly applicable to a wide range of fields, such as machine learning and signal processing. **Fig. 5b** shows the AMC circuit for solving the matrix equation $\boldsymbol{Ax} = \boldsymbol{b}$, where the matrix $\boldsymbol{A}$ is stored in the crosspoint array as conductance values, the known vector $-\boldsymbol{b}$ is provided to the array rows as input currents[22]. The rows are connected to the inverting-input terminals of OPAs, and the output of OPAs, which represents the equation solution $\boldsymbol{x}$, are fed back to array columns, forming a global feedback loop. At steady state, because of the virtual ground virtual short condition of OPA, the circuit must satisfy the equation $\boldsymbol{Ax} - \boldsymbol{b} = \boldsymbol{0}$, which implies the OPAs' outputs constitute the solution $\boldsymbol{x} = \boldsymbol{A}^{-1}\boldsymbol{b}$. In practice, the inputs are implemented as externally applied voltages through a fixed resistor, resulting in that both input and output are voltages. Before the input and after the output, DACs and ADCs can be connected to interface with the digital domain. The matrix inversion circuit can be considered equivalent to a recurrent neural network[66], with its circuit topology closely resembling that of a Hopfield network[67].

The critical role of the OPA as an integrator becomes evident when analyzing the circuit's temporal response[50,68]. By modeling the OPA as a single-pole system and applying the inverse Laplace transform, it is found that the circuit's dynamics are governed by a first-order differential matrix equation, namely $d\boldsymbol{x}/dt = \boldsymbol{U}^{-1}[\boldsymbol{Ax}(t) - \boldsymbol{b}]$, where $\boldsymbol{U}$ is a positive diagonal matrix whose elements are composed of summations of each rows of matrix $\boldsymbol{A}$. It is similar to $d\boldsymbol{x}/dt = \boldsymbol{Ax}(t) - \boldsymbol{b}$ that is implemented by analog CMOS circuits in **Fig. 4e**, except for the inclusion of the matrix $\boldsymbol{U}$ induced by KCL. Although no explicit capacitor is used for charge integration, the circuit characteristics dictate that the signal difference between the input terminals must be amplified with a gain consistent with the matrix inversion solution. Alternatively, a feedback capacitor can be explicitly included across the OPA, as in traditional configurations, though this would change the time constant[69,70]. According to the dynamics of the circuit described by the differential equation, it is required that the eigenvalues of matrix $\boldsymbol{U}^{-1}\boldsymbol{A}$ should all contain positive real parts, which is equivalent to that the poles of the system are located at the left half plane in the complex frequency plane[50]. Such a condition can be met by the positive definiteness of matrix $\boldsymbol{A}$, which is consistent with the case of analog CMOS circuit.

The same design principle has been applied to a series of AMC circuits. **Fig. 5c** and **5d** illustrate circuits for solving matrix eigenvector and generalized inverse, respectively. In **Fig. 5c**, the eigenvector circuit solves the matrix equation $\boldsymbol{Ax} = \lambda\boldsymbol{x}$, where $\boldsymbol{A}$ is a square matrix mapped onto the resistive memory array, $\lambda$ is a known eigenvalue (assumed to be positive) mapped onto the feedback conductance of TIAs, and $\boldsymbol{x}$ is the unknown eigenvector represented by the output voltages of the analog inverters[22]. An analog inverter consists of a TIA and a serial resistor. Since this circuit has no external input, it operates solely through a positive feedback mechanism, which gradually amplifies the circuit signals until the system reaches a steady state satisfying the eigenvector equation[71]. Due to the use of two sets of amplifiers, the circuit dynamics are described by a second-order differential matrix equation. For negative eigenvalues, the absolute value of the eigenvalue should instead be mapped onto the feedback conductance of the TIAs, and the analog inverters can then be removed. In this case, the circuit dynamics are described by a first-order differential matrix equation.

In **Fig. 5d**, the generalized inverse circuit solves the matrix equation $\boldsymbol{A}^{\mathrm{T}}\boldsymbol{A}\boldsymbol{x} = \boldsymbol{A}^{\mathrm{T}}\boldsymbol{b}$, which is a generalized equation of $\boldsymbol{A}\boldsymbol{x} = \boldsymbol{b}$. Here the matrix $\boldsymbol{A}$ is a tall matrix whose rows are more than columns, namely the linear system is overdetermined, thus the circuit gives the solution that minimizes least squares error $||\boldsymbol{A}\boldsymbol{x} - \boldsymbol{b}||$. The circuit consists of two resistive memory arrays that map two copies of matrix $\boldsymbol{A}$, a set of TIAs, and a set of positive feedback amplifiers[72]. The circuit should also be described by a second-order differential matrix equation, which is revealed by its transfer function in the figure[73]. According to the transfer function, the poles of the system can be determined. As a result, this generalized inverse circuit turns out to be always stable by the quadratic eigenvalue problem theory. It overcomes the limitations associated with specific matrix structures, providing a universal approach for solving matrix inversion problems, including those involving square matrices. This conclusion can be understood by observing that the matrix $\boldsymbol{A}^{\mathrm{T}}\boldsymbol{A}$ is always positive definite. By changing the input and output terminals, this circuit can also be used for solving inversion problems of broad matrix, which has more columns than rows. This generalized inverse concept has been very useful, and it has been extended to designing circuits for generalized least squares regression[74,75], and eigen-decomposition by solving all pairs of eigenvalue and eigenvector[76-78]. For all these circuits, processing of real-valued matrix containing negative elements can be performed using row (or) column splitting by analog inverters, or conductance compensation without analog inverters[79,80].

These circuits are primarily designed based on their static input-output characteristics, but their behavior is modeled using differential matrix equations derived from complex frequency analysis. A recent exception is the analog matrix computing (AMC) circuit used for solving sparse approximation problems, such as compressed sensing recovery[81] (**Fig. 5e**). This circuit is designed using the local competition algorithm (LCA), which itself is a first-order differential matrix equation[82]. Additionally, LCA incorporates an elementwise nonlinear function, typically the soft or hard threshold function[83], which suppresses the output to zero for specific input ranges around zero. The circuit still operates within the framework of solving $\boldsymbol{Ax} = \boldsymbol{b}$, which, however, is an underdetermined linear system (*i.e.*, $\boldsymbol{A}$ is a wide matrix with more columns than rows). Meanwhile, the solution is constrained to be sparse. In LCA, the most computationally intensive part is the matrix-matrix multiplication (acquisition of the Gram matrix)[84], which cannot be achieved with the single-array-based MVM operation. To address this, a matrix-matrix-vector multiplication (MMVM) circuit has been designed using the conductance compensation method, mapping two copies of matrix $\boldsymbol{A}$ in a single memory array. This approach is equivalent to performing two cascading MVM operations but occurs instantaneously without intermediate outputs[76,85], which is only feasible with resistive memory arrays that offer efficient summation, fan-out, and multiplication operations. Based on this MMVM circuit, a closed-loop circuit implementing LCA has been developed. Additional modules in the circuit include a set of analog subtractor-based soft threshold units, TIAs for current-to-voltage conversion, and analog inverters. Despite the three sets of amplifiers, the equation governing the circuit dynamics remains consistent with LCA after certain approximations. Experimental results demonstrate that compressed sensing recovery problems can be solved within a few microseconds, with the equilibrium state providing the solution.

**Solving matrix equations through matrix multiplication with resistive memory circuits**

Resistive memory arrays can also be used to accelerate MVM operations, which are the backbone of iterative algorithms for solving matrix equations. Given the fundamental importance of the matrix inversion problem $\boldsymbol{Ax} = \boldsymbol{b}$ in a wide range of applications, particularly in scientific computing, significant efforts have been made to speed up its solution using resistive memory arrays[20,21]. Various iterative algorithms, such as the conjugate gradient method, GMRES, and the Jacobi method[86-91], have been explored for this purpose, as illustrated in **Fig.**

**6a**. In these iterative algorithms, computations can generally be categorized into three fundamental types: (1) MVM operations, (2) vector operations, such as addition, subtraction, and dot product calculations, and (3) scalar operations, including multiplication and division for scaling. These operations are executed in an interleaved manner, forming an algorithmic feedback loop that is fundamentally different from the physical feedback loop employed in direct equation-solving circuits. Since MVM is typically the most computationally intensive operation, leveraging resistive memory arrays for MVM can significantly improve computational efficiency. However, the remaining operations are still performed in the conventional digital domain, which requires frequent data conversion between the analog and digital domains.

In principle, all these operations should be performed with high precision to ensure the reliable execution of iterative algorithms. However, because of the nonideal factors in the circuit, including imprecise writing of resistive memory devices, parasitic wiring resistances, and analog noises, a single MVM operation performed using a resistive memory array typically exhibits low precision, equivalent to only 4-6 bits[92-94]. Consequently, the overall precision of analog equation solving is limited and can be further degraded in ill-conditioned problems due to high condition numbers in matrix inversion. To enhance precision, two common approaches are employed, namely 1) using iterative refinement method in a high-precision digital computer to iteratively correct errors in the analog computation, gradually improving accuracy until the predefined precision is reached[20,86-88] (**Fig. 6b**), and 2) applying precision extension methods, such as bit slicing and analog compensation, to improve the precision of each MVM operation within the iterative algorithm[21,89-91] (**Fig. 6c**). Ultimately, achieving high-precision analog computing depends on the accurate representation of key numerical information[95], in this case matrix $\boldsymbol{A}$ and vector $\boldsymbol{b}$.

In the iterative refinement method, high-precision computation is ensured by a digital processor, which stores and processes the precise matrix $\boldsymbol{A}$ and vector $\boldsymbol{b}$. Meanwhile, the crosspoint resistive memory array holds an approximate version of $\boldsymbol{A}$ (denoted as $\widehat{\boldsymbol{A}}$), which is subject to imprecisions due to device variations and programming inaccuracies[20]. Each iteration begins with the digital processor performing a high-precision MVM operation using $\boldsymbol{A}$ and the current solution vector $\boldsymbol{x}$ (initialized to $\mathbf{0}$), yielding the residual vector $\boldsymbol{r}$. This residual vector is then sent to the low-precision MVM-based analog solver, which computes the incremental solution update $z$. The digital processor then integrates $z$ back into $\boldsymbol{x}$, progressively refining the solution (**Fig. 6b**). Since the analog solver provides an initial approximation of the solution, subsequent iterations rapidly converge toward the precise result. Typically, for large matrices (*e.g.*, thousands by thousands in size), the iterative refinement process achieves FP32 or FP64 precision within a few dozen cycles. Notably, the iterative cycles within the MVM-based analog solver itself are not included in this count. Because the analog solver already incorporates digital processing elements, it seamlessly integrates with the digital refinement process.

In the precision extension approach, high-precision analog MVM is achieved through techniques such as bit slicing, analog slicing or analog compensation[21,91,92]. These methods enable faithful implementation of iterative algorithms, but still maintaining an analog-digital hybrid architecture. As illustrated in **Fig. 6c**, precision extension is often combined with the block matrix method to handle large matrices that arise from discretizing continuous variables in scientific computing—matrices that exceed the size limitations of resistive memory arrays. To perform MVM efficiently, the original matrix is partitioned into multiple block matrices, and the corresponding vector is also divided accordingly[96,97]. Each block matrix contributes a partial MVM result, and all partial results must be aggregated to reconstruct the complete solution. To minimize error propagation from individual partial results, summation is preferably performed in the digital domain, requiring all partial outputs to be discretized via ADCs (**Fig. 6d**).

Each matrix (or block matrix) is initially high-precision, often represented in floating-point format, which must be converted to fixed-point for resistive memory implementation. The high-precision matrix is then bit-sliced into multiple lower-precision sub-matrices, each carrying different weight significance[98-100]. Similarly, the input vector can be bit-sliced into weighted sub-vectors. The MVM operation is performed sequentially for each pair of bit-sliced matrices and vectors, progressing from the most significant bits (MSBs) to the least significant bits (LSBs). As shown in **Fig. 6e**, if both the matrix and vector are split into two bit slices, four low-precision MVM operations are required using resistive memory arrays. The outputs are digitized by ADCs, appropriately shifted according to their bit significance, and summed via an adder tree, yielding a high-precision MVM result. Beyond bit slicing, alternative techniques such as analog compensation and analog slicing exploit the quasi-continuous tunability of resistive memory conductance to mitigate bit-width constraints, reducing hardware complexity and energy consumption[91,92,101]. However, the fundamental principle remains: decomposing high-precision matrices and vectors into multiple lower-precision components with distinct weights. During iterative algorithm execution, each cycle involves a high-precision MVM operation along with vector and scalar computations. The convergence behavior should ideally remain consistent with digital implementations; however, low-precision analog components are susceptible to noise, such as device and circuit $1/f$ noise. Since this noise is random and time-varying, it cannot be fully compensated, leading to fluctuations in the analog currents. These fluctuations are particularly detrimental to the MSB slices, where even minor deviations can significantly impact accuracy. Consequently, the convergence of the iterative solver is typically slower than that of digital implementations and stalls at a lower precision compared to an FP64 solver.

## Discussion

**Applications and the precision issue.** Modern analog computing is defined from both the perspectives of computing primitives and hardware implementations, driven by contemporary application demands, particularly in scientific computing. Traditionally, analog computing has focused on solving differential equations using analog circuits to model and analyze physical systems, with CMOS circuits serving as the primary means for direct analog solving. An alternative approach is to discretize differential equations into linear matrix equations, which can then be solved using either CMOS circuits or resistive memory arrays. In this case, the two approaches are functionally equivalent. For example, resistive memory array-based analog solvers have been used to solve differential equations, including Fourier's equation and Schrödinger's equation[22].

A critical challenge in modern analog computing is its inherently limited precision, which significantly constrains its application to solving differential and matrix equations. Whether analog computing is applied directly to differential equations or indirectly through matrix equations, precision limitations remain a fundamental bottleneck. Consequently, analog computing is often used to generate an approximate initial solution, which is subsequently refined using high-precision digital methods. This hybrid approach accelerates the overall computation process by leveraging the high speed of analog processing while compensating for its limited precision through digital correction. Conceptually, this approach is similar to the iterative refinement method illustrated in **Fig. 6b**. In addition, we propose that the low-precision MVM-based iterative algorithms shown in **Fig. 6a** can be directly replaced by the analog matrix inversion circuit in **Fig. 5b**. Furthermore, the high-precision MVM operations required in the iterative refinement process can also be implemented using analog computing. Owing to the distributive property of MVM, high-precision scientific computing can be achieved through analog MVM by employing precision-enhancement techniques such as bit slicing. By combining these approaches, it has been demonstrated that high-precision matrix-equation solving can be realized in a fully AMC-based framework[102].

Although resistive memory arrays have shown promise in analog computing, their application to directly solving differential equations remains underexplored. A key limitation is that resistive memory devices cannot naturally function as 2-input multipliers, making them less suitable for solving nonlinear differential equations directly. Nevertheless, recent research has demonstrated the potential of using a 1T1R cell as a 3-input multiplier[103,104], where two of the inputs are encoded as voltages, and the third is represented as nonvolatile conductance. This configuration opens up possibilities for implementing nonlinear solver designs. Beyond scientific computing, resistive memory array-based analog solvers have also been tested in other applications, such as machine learning[72,105-107] and signal processing[108-112]. These applications are naturally more tolerant of noise and imprecision, which alleviates the precision constraints typically associated with analog computing.

**Scalability.** The scalability of CMOS-based analog computing is fundamentally limited by the large area and high power consumption of CMOS analog operators. This challenge can be effectively mitigated by emerging resistive memory technologies, which have seen significant advancements in recent years. Thanks to their back-end-of-the-line (BEOL) compatibility, which enables substantial cell area reduction[113], high-density and reliable resistive memory arrays—such as RRAM and PCM—have become commercially viable, providing a strong foundation for large-scale equation solving. To date, large-scale matrix-vector multiplication (MVM) has been successfully demonstrated using analog computing with resistive memory, with crossbar arrays scaling up to 256×256[91]. However, extending this scalability to matrix equation solvers presents additional challenges. Unlike MVM, which operates in an open-loop configuration, solving matrix equations requires closed-loop feedback, which is difficult to implement within the current memory architectures. Additionally, the intrinsic fragility and limited endurance of resistive memory devices, as well as the increasing impact at large array sizes of parasitic elements, such as interconnection resistances and line capacitances, further constrain their application in large-scale equation solvers.

In principle, the scalability of analog equation-solving circuits is inherently limited by numerical stability concerns, particularly for ill-conditioned matrices. However, for specific classes of matrices, such as diagonally dominant ones, where the condition number constraint is relaxed, more aggressive scaling is possible. One potential approach to improve reliability and uniformity is to emulate resistive memory devices using a network of conductance-weighted integrated resistors controlled by SRAM-driven transmission gates[114-116]. This method combines the fast reprogrammability of SRAM memories with the ohmic behavior provided by integrated resistors to allow compatibility with the analog IMC work principle, alleviating high-nonlinearity and low-endurance effects of current-generation resistive memories. Using such an equivalent resistive memory array, experimental demonstrations have been extended to a 64×64 matrix size, although with significant area consumption for the memory array owing to the simultaneous presence of bulky SRAM and integrated resistors[114]. In addition, multi-core systems[117] can be envisioned relying on problem decomposition techniques and block matrix algorithms, such as BlockAMC[118]. These algorithms extend the principles of block matrix methods used in large-scale MVM but integrate both block matrix multiplication and block matrix inversion to enable efficient large-scale matrix equation solving. With this approach, the effective addressable matrix size has been extended to 512×512, which is substantial for many applications, including solving discretized differential equations in scientific computing and second-order training of large-scale neural networks.

**Relationship with in-memory computing.** Due to the extensive use of resistive memory devices, modern analog computing is closely linked to the concept of in-memory computing. In particular, resistive memory array-based MVM acceleration has been widely studied within the in-memory computing framework, where neural network workloads are a primary focus[119,120]. In this context, analog MVM serves as a critical bridge between equation solving in analog computing and MAC acceleration in in-memory computing. Additionally, some studies have explored the acceleration of MAC operations specifically for solving discretized differential equations[121,122]. On the other hand, CMOS analog chips, such as the one shown in **Fig. 4a**, can be regarded as a form of near-memory computing, as they integrate SRAM directly with analog operators on the same chip. In conventional von Neumann architectures, data movement—not computation—dominates both latency and energy consumption. By eliminating or reducing data transfer overhead during computation, in-memory and near-memory computing architectures can significantly enhance computational efficiency at the system level. While this energetic advantage can be offset by the more intensive analog circuitry requirements, including energy-hungry ADCs and OPAs, capacitive memory-based systems provide a promising pathway for energy reduction[123].

**Computational complexity.** Complexity analysis and optimization are fundamental to theoretical computer science, originally developed within the domain of digital computing. While the complexity of analog computing has been discussed in the past, such discussions were primarily limited to differential equation-solving circuits, where the impact of resistive and capacitive elements was analyzed to understand how increasing circuit scale affects the time constant[124]. More recently, efforts have been made to analyze the computational complexity of analog matrix computing[19,50,71,73]. Compared to differential equation solvers, analog matrix equation-solving circuits offer a clearer comparison to digital computing, as they can be interpreted as continuous-time iterative algorithms. Ideally, MVM in a resistive memory crossbar array should be computed instantaneously, at the speed of electromagnetic wave propagation in the dielectric medium. However, in practical implementations, the presence of readout circuitry (*e.g.*, TIAs) and parasitic capacitance introduces a time constant that is proportional to the maximum conductance summation in the array but remains independent of the matrix size[19]. This property also extends to matrix equation-solving circuits.

In an ideal case where only resistive devices and single-pole OPAs are considered, analytical results indicate that the computation time of these circuits is determined by the eigenvalues of a matrix associated with the original problem rather than directly scaling with matrix size[50]. This eigenvalue dependence in time complexity is also observed in digital iterative algorithms, in addition to their dependency on matrix size. The computational speed advantage of analog computing arises from the high parallelism of resistive crossbar arrays, which inherently integrate massive numbers of analog multipliers, adders, and fan-out operations, further enhanced by parallel OPAs. While the theoretical complexity advantage of analog computing over digital computing appears significant, these conclusions are derived under idealized conditions and primarily apply to single-array implementations. For arbitrarily large-scale matrix problems with real-world hardware constraints, factors such as interconnect delays, noise, and precision limitations may reduce this advantage, necessitating further investigation into practical scalability and performance.

To conclude, modern analog computing has achieved substantial advancements in recent years, encompassing both computational primitives and hardware implementations. These innovations enable significant acceleration for workloads in artificial intelligence, scientific computing, and interdisciplinary domains. In turn, driven by the escalating computational demands of these applications, analog computing is poised for renewed growth—propelling CMOS analog circuit technology toward more advanced nodes and fostering novel resistive memory architectures. While resistive memory has already reached milestones in traditional storage and in-memory computing, its application in analog equation-solving systems represents a promising new frontier, capable of addressing increasingly complex tasks such as solving differential equations in generative artificial intelligence (*e.g.*, diffusion models)[125]. By leveraging the structural simplicity and spatial regularity of resistive memory, challenges of scalability and precision in analog equation-solving have also been partially mitigated, through hybrid analog–digital strategies such as iterative refinement and bit slicing. Given the relentless pursuit of high-performance computing and the foundational role of differential equations and matrix computations, we hope this paper serves as a catalyst for the continued development and deployment of modern analog computing.

## Acknowledgements

This work was supported by National Natural Science Foundation of China (62572011), Beijing Natural Science Foundation (4252016) and the 111 Project (B18001).

## Author Contributions

All authors contributed to the preparation of the manuscript.

## Competing interests

The authors declare no competing interests.

## Additional information

**Correspondence** should be addressed to Zhong Sun.

## FIGURE CAPTIONS

**Fig. 1. Analog computing primitives**, including solving ordinary differential equation, solving matrix equation, and performing matrix multiplication, each of which are all is implementable with analog hardware. The algorithmic approach in digital computing is also illustrated for comparison. Notably, algorithms are partially involved in analog computing as well, to decompose the computing tasks. For example, a differential equation can be discretized in the spatial and /or temporal domains and transformed into a matrix equation (a system of linear equations). Similarly, solving a matrix equation can be decomposed into a sequence of matrix multiplications and other scalar/vector operations within an iterative algorithm.

**Fig. 2. Circuit schematics of the three computing primitives. a**, Basic analog operators. Circuit schematics for **b**, matrix-vector multiplication, **c**, solving an ordinary differential equation, and **d**, solving a matrix equation. In these schematics, the constant input operators have been omitted for clarity and instead are represented by a floating arrow.

**Fig. 3. Different hardware implementations of analog operators.** In the era of discrete components, analog adder and integrator are usually implemented with OPAs, working in the voltage domain. In the era of IC, integrated CMOS amplifiers, *e.g.*, Gilbert cell and OTA, are used for analog multiplier and integrator. Gilbert cell may work in the pure voltage mode, multiplying two differential voltages and outputting another differential voltage. The output can also be considered as the differential current through the two resistors, and one input can be considered as the different current through the bottom differential pair, as indicated by the arrows in the schematic. The involvement of current is beneficial for the implementation of an analog adder. Some other configurations of Gilbert cell may work in the pure current mode, by leveraging the trans-linear relationship of currents through the transistors working in the weak inversion region. The OTA integrator may work in either voltage or current domain, upon the placement of the capacitor in the output or input terminal of the OTA, respectively. The dashed rectangle highlights that a crossbar array of resistive memory is an integration of analog multipliers, adders, and fan-out modules, when the circuit is working in hybrid mode.

**Fig. 4. CMOS analog chip for solving differential equations. a**, Chip architecture, from top to bottom are located analog, mixed-signal and digital blocks, to keep interference of the system low. **b**, Configuration of the chip for solving the Van der Pol equation. As an example, a set of initial conditions are provided. **c**, Analog solution to the Van der Pol equation from the chip. The temporal output should be continuously sampled. **d**, Configuration of the chip for solving the matrix equation $\boldsymbol{Ax} = \boldsymbol{b}$, where $\boldsymbol{A}$ is a 2×2 positive definite matrix. **e**, Analog solution to the matrix equation, where only the steady-state output needs be sampled.

**Fig. 5. Resistive memory circuits for solving matrix equations.** **a**, Schematic of analog conductance characteristics of a resistive memory cell, which could be a passive 1R device, or adopt the 1T1R structure. In the latter case, if the transistor is only used to control the accessibility of the cell, it can be effectively integrated into a crossbar. By applying appropriate voltage stimulus, the memory cell can be *set* to higher conductance state or *reset* to lower conductance state, in an analog (or multi-level) manner. Resistive memory array-based analog circuit for solving **b**, Matrix inversion equation $\boldsymbol{Ax} = \boldsymbol{b}$, **c**, Matrix eigenvector equation $\boldsymbol{Ax} = \lambda\boldsymbol{x}$, **d**, Generalized matrix inverse equation $\boldsymbol{A}^{\mathrm{T}}\boldsymbol{Ax} = \boldsymbol{A}^{\mathrm{T}}\boldsymbol{b}$, and **e**, Sparse approximation problem with LCA algorithm. In **b**-**d**, the time-domain dynamic equations or complex frequency-domain transfer function are also included. $f$ is the GBWP of the OPAs in use, assuming both TIAs and analog inverters adopt the same OPA. In **b**, $\boldsymbol{U}$ is a diagonal matrix whose elements are dictated by the summations of each row in matrix $\boldsymbol{A}$. In **c**, $\boldsymbol{U}$ is again a diagonal matrix, and $\boldsymbol{I}$ is the $n \times n$ identity matrix. In **d**, $s$ is the complex frequency variable, $\boldsymbol{U}_n$ ($\boldsymbol{U}_m$) is a diagonal matrix whose elements are dictated by the summations of each row (column) in matrix $\boldsymbol{A}$. In **e**, $\tau$ is a time constant, $\boldsymbol{u}$ is the internal state vector of the neurons, $\phi$ is the threshold of the neuron. Concom is a column of resistive memory devices for conductance compensation, to let the conductance summation along each row in the array is equal.

**Fig. 6. MVM-based approaches for solving $\boldsymbol{Ax} = \boldsymbol{b}$.** **a**, Iterative algorithms for solving the matrix equation, with the core MVM operation implemented with a crosspoint resistive memory array. A single crosspoint array can only deliver approximate results. **b**, Iterative refinement with a high-precision digital computer. **c**, Bit slicing and block matrix methods for implementing large-scale, high-precision matrix $\boldsymbol{A}$. **d**, Partial MVM results with block matrices and input sub-vectors. Each block matrix is implemented by a crosspoint array, and the partial MVM results should be accumulated in the digital domain. **e**, Partial MVM results with bit-sliced matrix and vector of MSBs

and LSBs. Because the sliced bits have different weight significances, the resulting partial MVM results should be processed with shift-and-add in the digital domain, to recover the high-precision result.

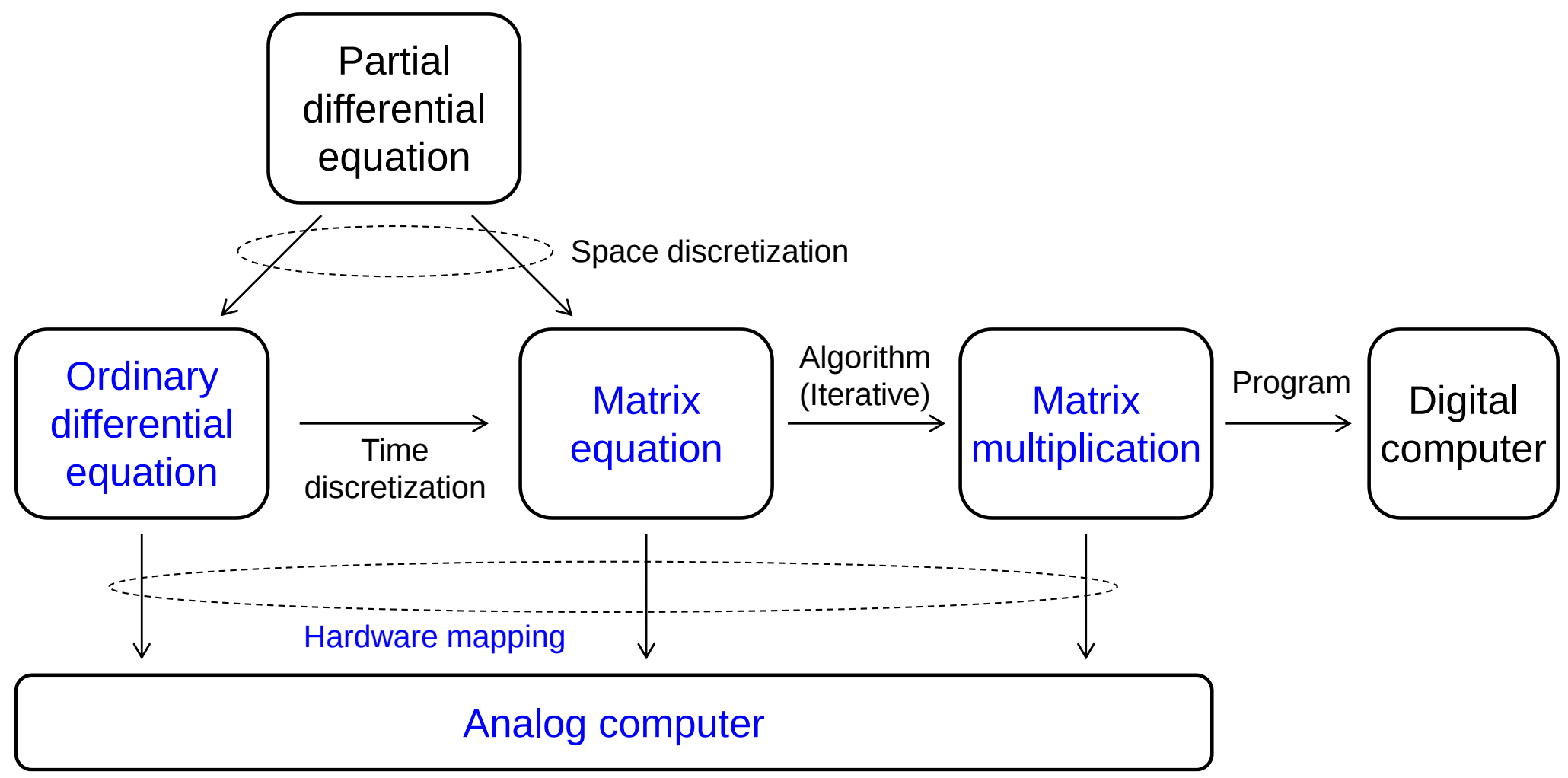
Partial differential equation
Space discretization
Ordinary differential equation
Time discretization
Matrix equation
Algorithm (Iterative)
Matrix multiplication
Program
Digital computer
Hardware mapping
Analog computer

**Figure 1.**

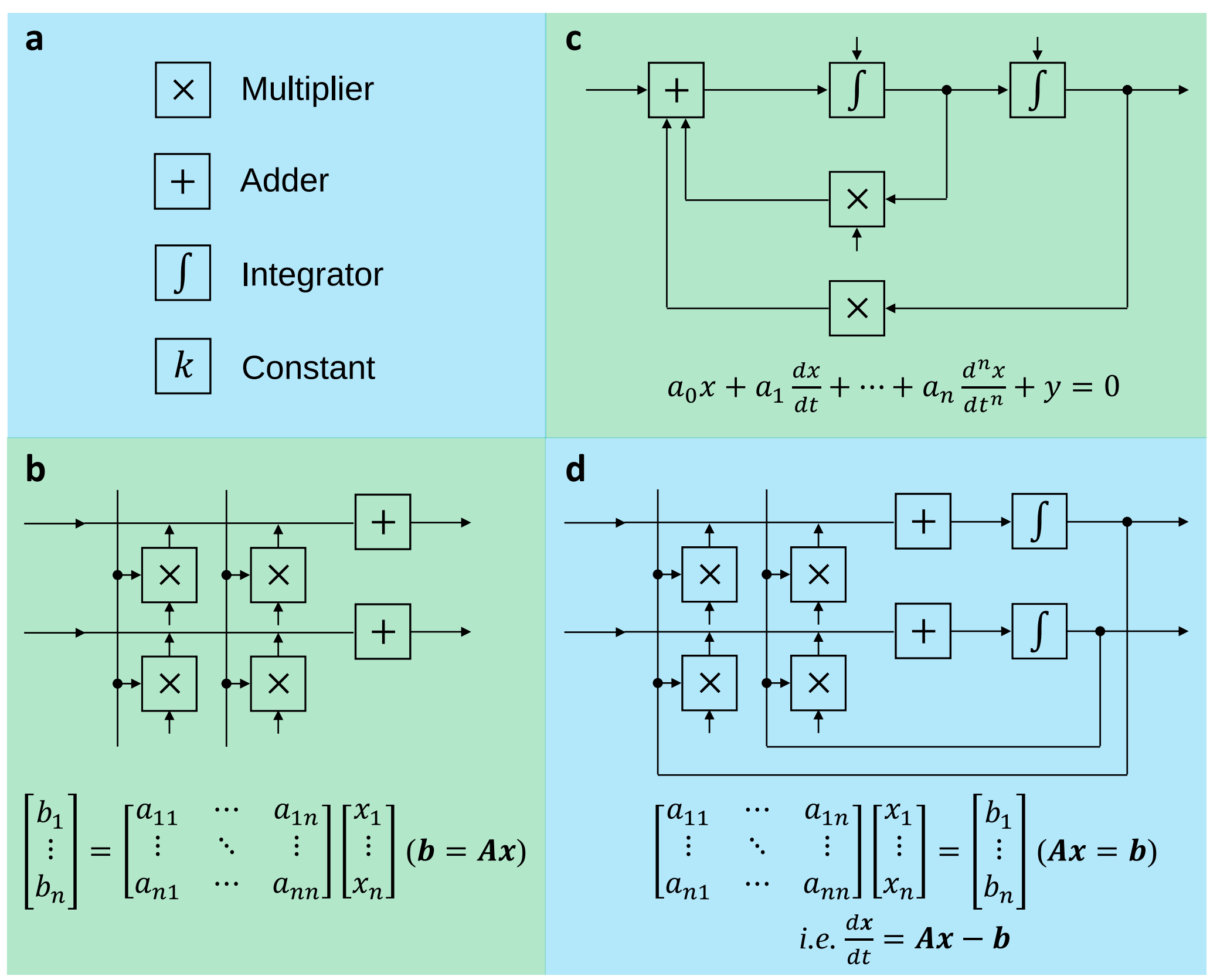
a
Multiplier
Adder
Integrator
$k$
Constant
b
$\begin{bmatrix} b_1 \\ \vdots \\ b_n \end{bmatrix} = \begin{bmatrix} a_{11} & \cdots & a_{1n} \\ \vdots & \ddots & \vdots \\ a_{n1} & \cdots & a_{nn} \end{bmatrix} \begin{bmatrix} x_1 \\ \vdots \\ x_n \end{bmatrix}$ $(\boldsymbol{b} = \boldsymbol{Ax})$
c
$a_0 x + a_1 \frac{dx}{dt} + \cdots + a_n \frac{d^n x}{dt^n} + y = 0$
d
$\begin{bmatrix} a_{11} & \cdots & a_{1n} \\ \vdots & \ddots & \vdots \\ a_{n1} & \cdots & a_{nn} \end{bmatrix} \begin{bmatrix} x_1 \\ \vdots \\ x_n \end{bmatrix} = \begin{bmatrix} b_1 \\ \vdots \\ b_n \end{bmatrix}$ $(\boldsymbol{Ax} = \boldsymbol{b})$
i.e. $\frac{dx}{dt} = \boldsymbol{Ax} - \boldsymbol{b}$

**Figure 2.**

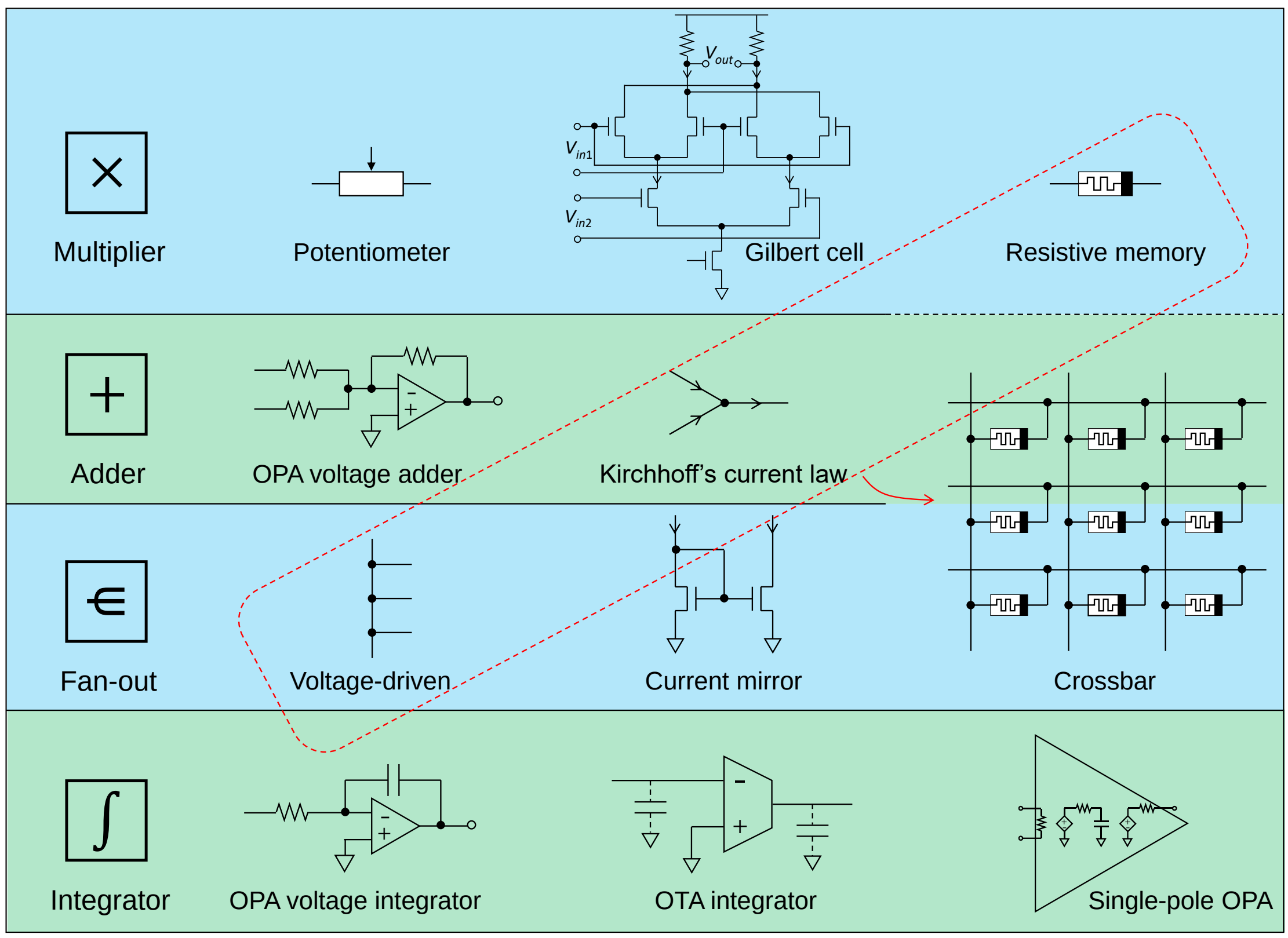

Multiplier
Potentiometer
$V_{out}$
$V_{in1}$
$V_{in2}$
Gilbert cell
Resistive memory
Adder
OPA voltage adder
Kirchhoff's current law
Fan-out
Voltage-driven
Current mirror
Crossbar
Integrator
OPA voltage integrator
OTA integrator
Single-pole OPA

**Figure 3.**

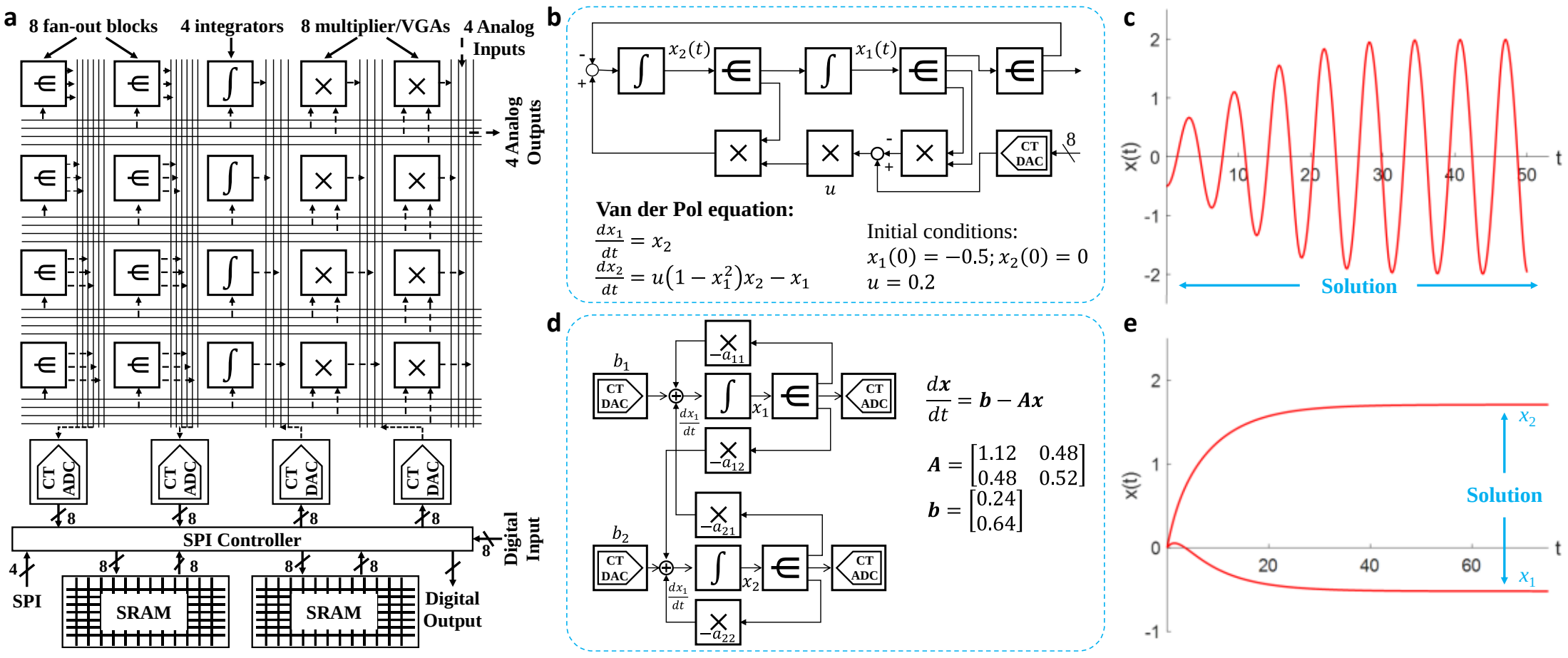

a
8 fan-out blocks
4 integrators
8 multiplier/VGAs
4 Analog Inputs
4 Analog Outputs
CT ADC
CT DAC
SPI Controller
Digital Input
SPI
SRAM
Digital Output
b
Van der Pol equation:
Initial conditions:
$x_1(0) = -0.5; x_2(0) = 0$
$u = 0.2$
c
x(t)
Solution
d
$\frac{dx}{dt} = b - Ax$
e
Solution

**Figure 4.**

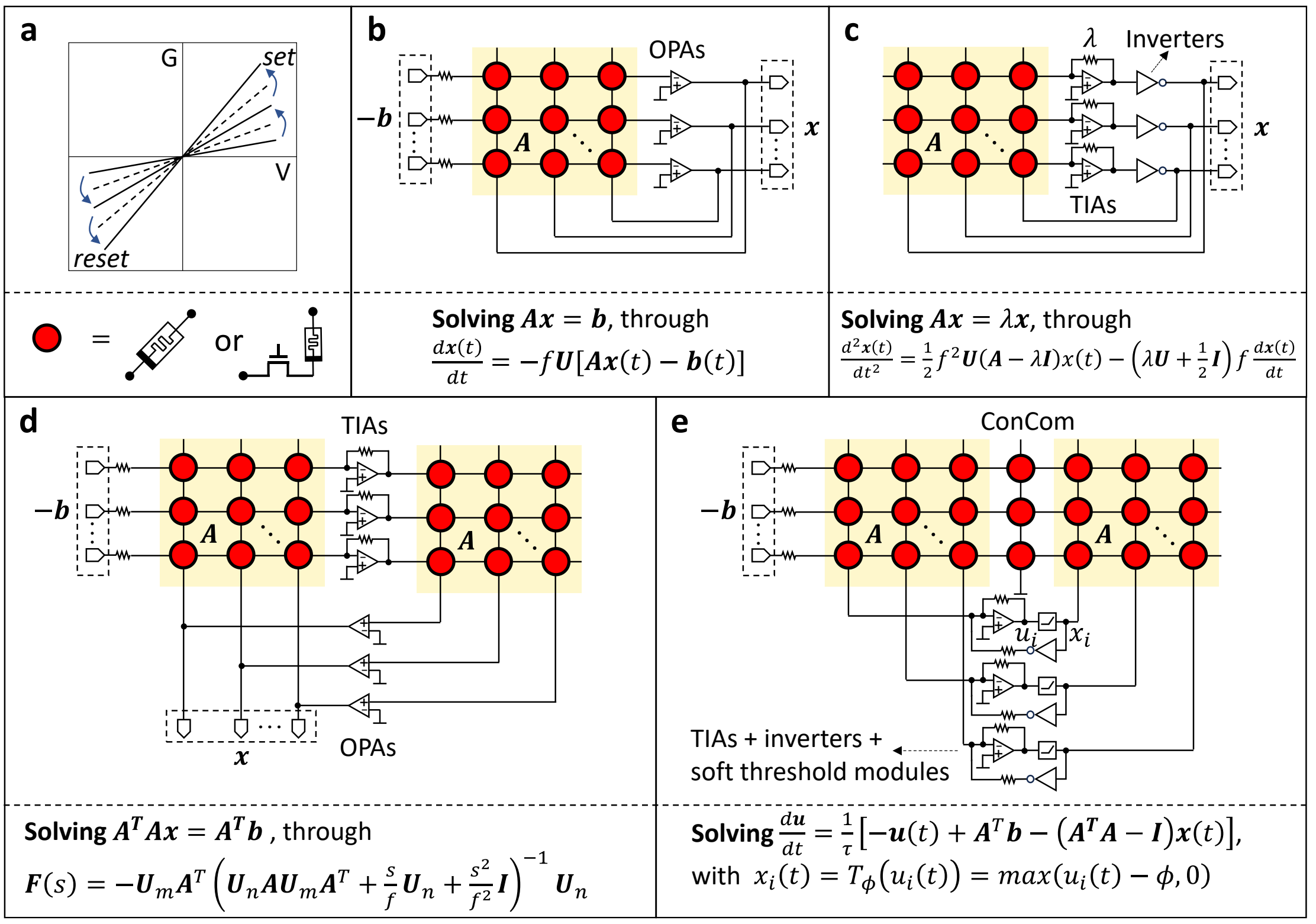
a
G
set
V
reset
= or
b
OPAs
−b
A
x
Solving $Ax = b$, through
$\frac{dx(t)}{dt} = -fU[Ax(t) - b(t)]$
c
λ Inverters
A
x
TIAs
Solving $Ax = \lambda x$, through
$\frac{d^2x(t)}{dt^2} = \frac{1}{2}f^2U(A - \lambda I)x(t) - \left(\lambda U + \frac{1}{2}I\right)f\frac{dx(t)}{dt}$
d
TIAs
−b
A
A
x
OPAs
Solving $A^TAx = A^Tb$, through
$F(s) = -U_mA^T\left(U_nAU_mA^T + \frac{s}{f}U_n + \frac{s^2}{f^2}I\right)^{-1}U_n$
e
ConCom
−b
A
A
$u_i$
$x_i$
TIAs + inverters +
soft threshold modules
Solving $\frac{du}{dt} = \frac{1}{\tau}[-u(t) + A^Tb - (A^TA - I)x(t)]$,
with $x_i(t) = T_\phi(u_i(t)) = max(u_i(t) - \phi, 0)$

**Figure 5.**

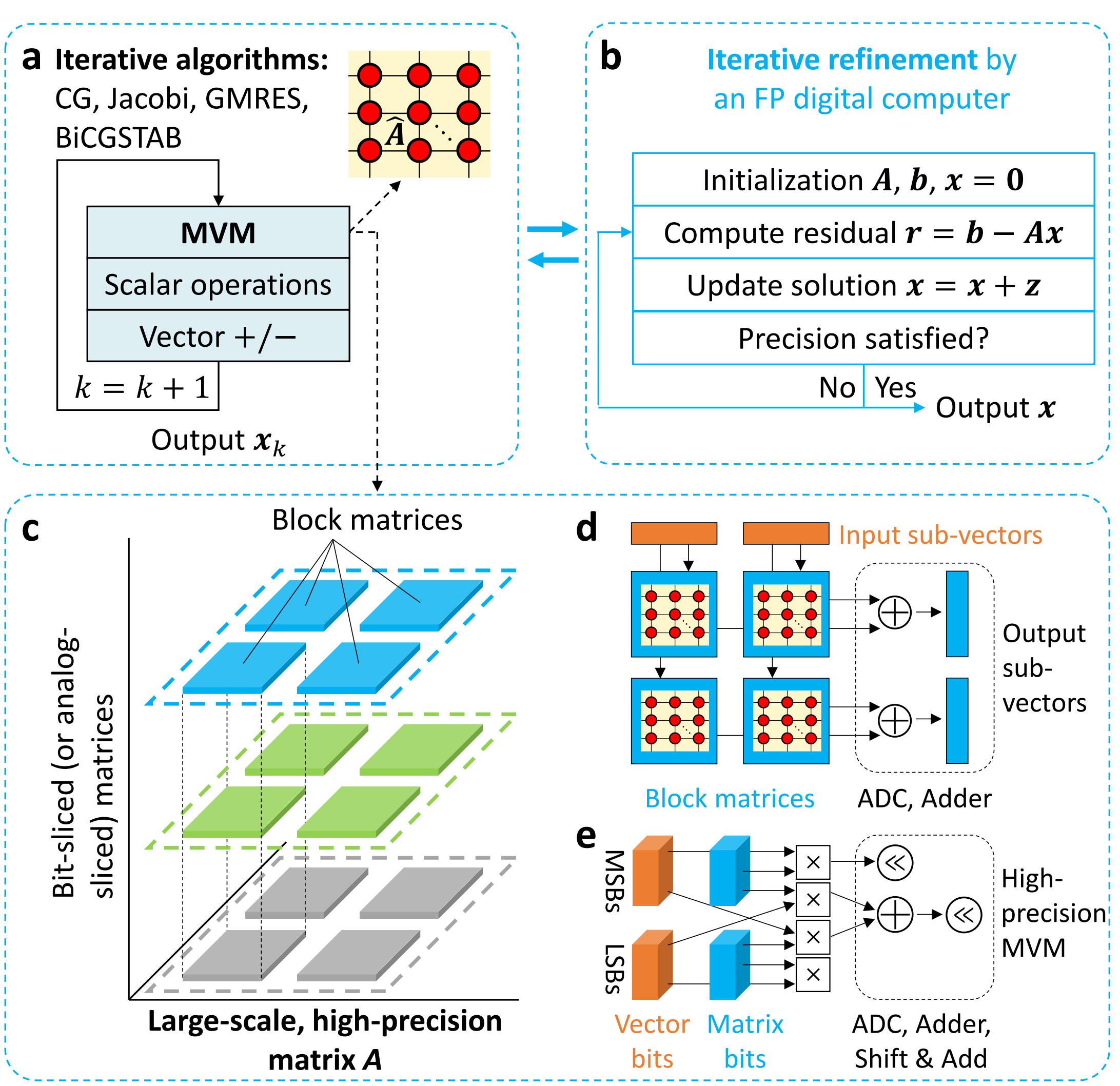
a Iterative algorithms:
CG, Jacobi, GMRES, BiCGSTAB
MVM
Scalar operations
Vector +/−
$k = k + 1$
Output $\boldsymbol{x}_k$
b Iterative refinement by an FP digital computer
Initialization $\boldsymbol{A}$, $\boldsymbol{b}$, $\boldsymbol{x} = \boldsymbol{0}$
Compute residual $\boldsymbol{r} = \boldsymbol{b} - \boldsymbol{A}\boldsymbol{x}$
Update solution $\boldsymbol{x} = \boldsymbol{x} + \boldsymbol{z}$
Precision satisfied?
No
Yes
Output $\boldsymbol{x}$
c
Block matrices
Bit-sliced (or analog-sliced) matrices
Large-scale, high-precision matrix $\boldsymbol{A}$
d
Input sub-vectors
Output sub-vectors
Block matrices
ADC, Adder
e
MSBs
LSBs
Vector bits
Matrix bits
ADC, Adder, Shift & Add
High-precision MVM

**Figure 6.**